\documentclass[twoside]{dis04}
\def\runauthor{R. Enberg et al.\ }
\def\shorttitle{Diffractive photoproduction of vector mesons at large momentum transfer}
\begin{document}

\title{Diffractive photoproduction of vector mesons at large momentum transfer \\}

\author{R. Enberg}
\address{Centre de Physique Th{\'e}orique, {\'E}cole Polytechnique, \\
91128 Palaiseau Cedex, France}
\author{ J. Forshaw, G. Poludniowski}
\address{Department of Physics \& Astronomy, University of Manchester, \\
Manchester M13 9PL, UK
} 
\author{\underline{L. Motyka}}
\address{Institute of Physics, Jagellonian University, \\
Reymonta 4, 30-059 
Krak\'{o}w, Poland}

\maketitle

\abstracts{
Diffractive photoproduction of $\rho$, $\phi$ and $J/\psi$ was studied
in the BFKL approach to hard colour singlet exchange. 
Differential cross sections, the energy dependence and spin density matrix elements were calculated and compared to data from HERA. The overall description of data is reasonably good, except of the single flip amplitude which has the 
wrong sign. Importance of chiral odd components of the photon is stressed. 
}

\section{Introduction}

Diffractive photoproduction of vector mesons (VM) off a proton target at
large momentum transfer is a process observed at HERA at rather high
rates~\cite{ZEUS,H1}. 
In this process the proton, typically, dissociates forming a diffractive
system and vector mesons are observed via their decay channels. The
recorded statistics is quite high for $\rho$, $\phi$ and $J/\psi$
production up to momentum transfer of $|t| \sim 20$~GeV$^2$. The most
interesting observables that are measured are the differential
cross-section $d\sigma / dt$, its dependence on the $\gamma p$ collision
energy $W$ and the spin density matrix elements $r^{04} _{ij}$. The
determination of the latter is possible from the angular distribution
of the decay products of the mesons. The spin density matrix elements are
governed by the photoproduction amplitudes of a polarised vector meson by a
polarised quasi-real photon.

It follows from a phenomenological rule of the $s$-channel helicity
conservation (SCHC) that the meson should have the same polarisation as the
incoming photon. Indeed, this option dominates but the data for $\rho$
and $\phi$ exhibit some deviations from the SCHC scenario. It was determined
that the the amplitude with a single helicity flip $M_{+0}$ (that is a
photon with helicity +1 going into longitudinally polarised meson) and
double helicity flip, $M_{+-}$, measure about 10\% and 20\% of the leading,
helicity conserving amplitude, $M_{++}$, respectively. Thus, the simplest
scheme of SCHC is not sufficient and it is worth performing a QCD analysis
of the process. The requirement of obtaining a good simultaneous
description of the shapes and magnitudes of cross sections and the spin
density matrix elements is rather stringent and discriminates between
various QCD-based models.

In perturbative QCD vector meson photoproduction at high energies is
mediated by an exchange of a gluonic system in a colour singlet state. In
the leading order approximation the system is just two elementary,
non-interacting gluons. A detailed calculation \cite{IKSS} for light vector
mesons based on this assumption implied that the single flip amplitude
should be leading at sufficiently large momentum transfer, and the total
cross section should have an approximate power-like behaviour, $d\sigma /
dt \sim 1/|t|^3$. The shape of the cross section agrees with the data, but
the prediction about the leading amplitude is not correct. Even worse,
according to the model the actually leading $M_{++}$ amplitude would give
$d\sigma/dt \sim 1/|t|^4$. In search of the source of the discrepancy, an
important idea was put forward \cite{IKSS} that a non-perturbative component
of the photon wave function, related to chiral odd quark operators, makes
important contribution to the production amplitudes. In a perturbative
approach, the current light quark mass is used, which is negligibly small
and may be set to zero (this is the reason that the $M_{++}$
amplitude is naively expected to be suppressed in the light meson case). 
The QCD vacuum, however, is a medium which breaks
the chiral symmetry, the phenomenon responsible, for instance, for the
generation of the constituent mass of quarks. Indeed, the chiral odd
no-flip amplitude, $M_{++} ^{\mathrm odd}$, was found to be the largest one
at a moderate momentum transfer $|t|<20$~GeV$^2$. Still, a good
quantitative description of the bulk of data was not reached.

On the other hand, the helicity averaged differential cross sections for
the VM photoproduction are well described by the leading logarithmic BFKL
formalism with non-relativistic wave functions, both for $J/\psi$ and for
the light vector mesons \cite{FP}. Remarkably, in the non-relativistic
picture the whole contribution to the cross section is given by the
chiral odd no-flip amplitude and the constituent quark masses naturally
enter the calculation. A main drawback of this approach is an inability to
describe deviations from the SCHC and in order to improve it one needs to
go beyond the non-relativistic approximation. Thus, the main goal of our
analysis \cite{EMP,EFMP,EFMP2} 
was to employ the non-forward BFKL equation \cite{BFKL} to describe the hard
colour singlet exchange and combine it with a QCD guided description of the
meson wave functions.

\section{Formalism}

The BFKL equation in the leading logarithmic approximation describes the
evolution of the diffractive scattering amplitude with the rising rapidity
distance $Y$ between the colliding objects. Perturbative QCD corrections to
the simple two gluon exchange have leading pieces $\sim (\alpha_s Y)^n$, and
in spite of $\alpha_s$ being small, the higher order terms cannot be
neglected. Thus, the BFKL equation resums ladder diagrams, with reggeised
gluons along the ladder. The equation is an integral equation in the
transverse momentum of the gluons. The integral kernel exhibits the global
conformal invariance, when expressed in the complex representation of gluon
transverse positions, and due to that symmetry the Eigenfunctions of the
BFKL integral kernel $E_{n,\nu}$ may be found analytically. In this
representation, the BFKL amplitude may be written in a compact form, for any
momentum transfer $\vec{q}$ as an infinite sum over all conformal spins
$n$,
\begin{equation}
M(\vec{q},Y) = \sum_{n=-\infty} ^{\infty} 
\int d\nu\; {( \nu^2 + n^2/4)\exp[\alpha_s Y \chi_n(\nu)]  \over  [\nu^2 + (n-1)^2/4)^2][\nu^2 + (n+1)^2/4]}\; (E_{n,\nu}|\Phi_1)(\Phi_2|E_{n,\nu})
\label{master}
\end{equation}
The Eigenvalues of the BFKL kernel 
$\chi_n(\nu)=4 \mathrm{Re}\,\left[\psi(1) - \psi(1/2 + |n|/2 + i\nu)\right]$ 
govern the dependence on
the rapidity, and for $Y\gg 1$ the conformal spin $n=0$ yields the leading
contribution, giving the amplitude which grows with rapidity, $M \sim
\exp(12Y\alpha_s \, \ln 2 / \pi)$. Impact factors $\Phi_i$ are amplitudes
for transition of the projectile and the target into their final states,
e.g.\ a $\gamma \to V$ and $p \to X$ transitions in the VM photoproduction.
The impact factors in Eq.\ \ref{master} have been, symbolically, projected
on the the BFKL Eigenfunctions.

The helicity dependent impact factors, $\Phi(\gamma \to V)$, may be 
calculated in perturbative QCD under some assumptions about the wave 
functions of the polarised vector meson and of the polarised photon. 
The hard colour singlet exchange is a short distance
process, thus the short distance expansion for the vector meson wave
function is a natural starting point. Following \cite{BBKT,BB,IKSS} we used
QCD distribution amplitudes up to twist~3, taking into account both chiral
even and chiral odd ones. We chose to use the perturbative expression for
the photon wave function, with a constituent quark mass $m_q$. The mass is
an effective parameter here and it plays a dual r\^{o}le: it sets the
magnitude of the chiral odd pieces and it provides an infrared cutoff for
the size of hadronic system that the photon fluctuates into. In both cases,
the actual constituent light quark mass $m_q \simeq 0.3-0.4$~GeV is a sensible
choice. Sensitivity of the amplitudes to the value of the infrared cutoff
gives an estimate the validity of the short distance expansion.

We treated the impact factor describing the diffractive proton dissociation
in the standard way. At large momentum transfer the BFKL pomeron couples
predominantly to individual partons in an incoherent way. Therefore the
cross section may be factorised into partonic cross sections and partonic
densities. The issue of how to define the BFKL impact factor for a 
colourful object was studied before \cite{MT,MMR}.

\section{Results}

\begin{figure}[!thb]
\vspace*{6.1cm}
\begin{center}
\includegraphics{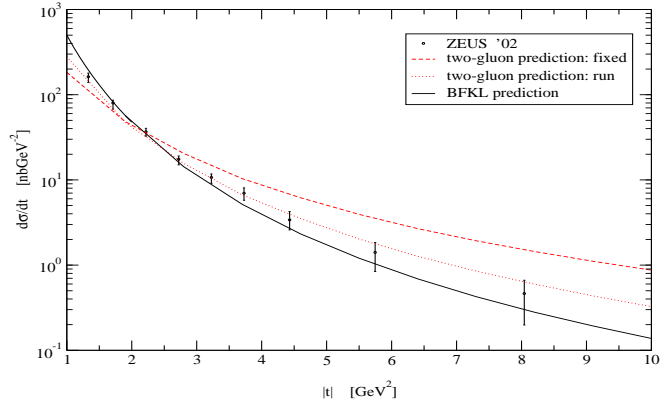}
\label{dsdt}
\vspace{5mm}
\caption[*]{$d\sigma / dt$ for diffractive $\rho$ photoproduction: ZEUS data and theory prediction in the two gluon approximation with fixed $\alpha_s$ (dashed line),  running $\alpha_s$ (dotted line) and the BFKL results (continuous line). }
\end{center}
\end{figure}

\begin{figure}[!thb]
\vspace*{6.0cm}
\begin{center}
\includegraphics{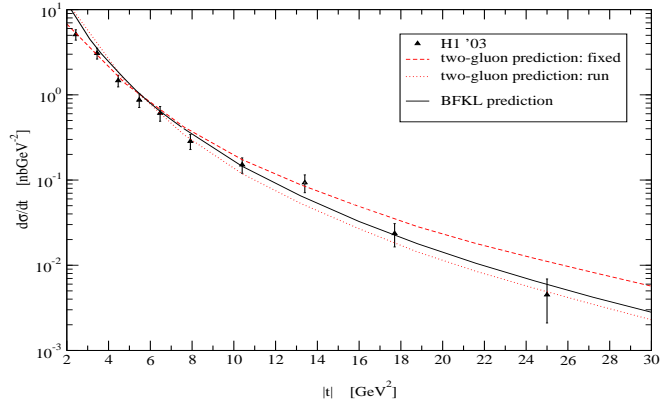}
\label{dsdtpsi}
\vspace{5mm}
\caption[*]{$d\sigma / dt$ for diffractive $J/\psi$ photoproduction: 
H1 data and theory prediction in the two gluon approximation with 
fixed $\alpha_s$ (dashed line),  running $\alpha_s$ (dotted line) 
and the BFKL results (continuous line). }
\end{center}
\end{figure}

The essential parameters of our analysis were the strong coupling constant
that scales the overall normalisation ($\alpha_s ^{IF}$), the strong
coupling constant that governs the BFKL rapidity dependence 
($\alpha_s ^{BFKL}$) 
and the value of constituent quark mass. The different values of
$\alpha_s^{IF}$ and $\alpha_s ^{BFKL}$ reflect the fact that non-leading QCD
corrections to the BFKL intercept and to the impact factors may be very
different. For reference, we chose to set the constituent quark mass to a
half of the meson mass. We included contribution to the scattering amplitudes
from all the conformal spins. The parton level BFKL amplitudes were 
convoluted with the parton densities, respecting the experimental cuts.

We determined the BFKL evolution of all the
independent helicity amplitudes for $\rho$, $\phi$ and $J/\psi$
photoproduction. All end-point infra-red divergencies which were found in
the two-gluon approximations disappear for rapidities $Y>0$ which justifies
the perturbative approach. We observed that the BFKL enhancement of the
chiral odd no-flip amplitude $M_{++}$ is the strongest, and that this part
of the amplitude gives the dominant contribution to the cross section.
The relative significance of the single flip amplitude turned out to be much
smaller than it was in the two-gluon exchange approximation, see 
Fig.~3.

{\bf Cross sections.} The hadronic level cross sections were calculated 
using partonic cross sections for all possible polarisations of the photon 
and of the meson. For light vector mesons the shape of differential cross 
section is well reproduced by the BFKL curve (see Fig.~1), 
the two-gluon 
exchange approximation gives an equally good fit for the running $\alpha_s$
and does somewhat worse for the fixed coupling. Hence, after inclusion 
of the chiral odd component of the photon, the dominance of the no-flip 
amplitude was found not to contradict the $\sim |t|^{-3}$ dependence 
of $d\sigma/ dt$. 
The results look rather similar for the $\phi$ photoproduction.
In the $J/\psi$ case (see Fig.~2), we found that a good fit 
of $d\sigma/ dt$ may be obtained both in the BFKL approach and in the
two gluon approximation if the QCD distribution amplitudes are used 
to describe the $J/\psi$ wave function \cite{EFMP,EFMP2}. 
The $t$-shape is stable against the treatment 
of the higher order QCD corrections also 
when the non-relativistic approximation is employed \cite{FP,EMP}.

\begin{figure}[!thb]
\vspace*{6.5cm}
\begin{center}
\includegraphics{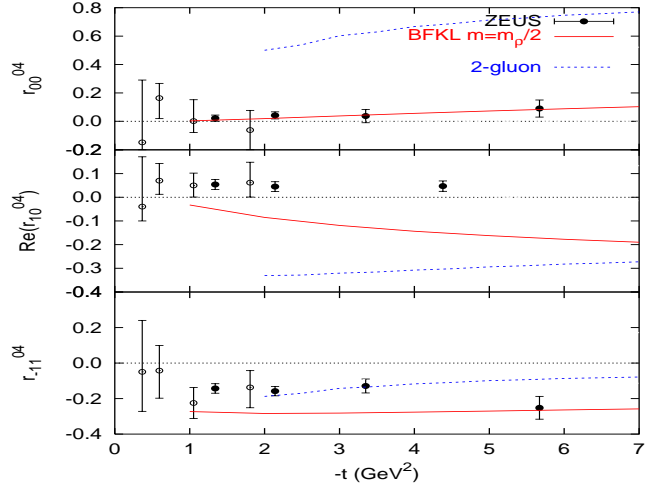}\vspace{5mm}
\caption[*]{ Spin density matrix elements: ZEUS data compared with the 
results of the BFKL calculation, and the two-gluon exchange approximation 
as functions of the momentum transfer.}
\end{center}
\label{rmat}
\end{figure}

{\bf Spin density matrix}. The angular dependence of the photoproduced
vector meson decay products is characterised by three elements of the spin
density matrix:
$r^{04}_{00} \sim |M_{+0}|^2$,
$r^{04}_{10} \sim {\mathrm Re}\, [M_{+0}^* (M_{++}-M_{+-})]$ and
$r^{04}_{1\, -1} \sim {\mathrm Re}\, [(M_{++} M_{+-}^*)]$.
In Fig.~3 we show the spin density matrix data compared with the
results of the BFKL calculation \cite{EFMP,EFMP2}. 
It is clear, that $r^{04}_{00}$ comes out
right, $r^{04}_{10}$ has the wrong sign and $r^{04}_{1\, -1}$ is too negative.
This means that the sign of the single flip amplitude is the
most serious problem. Let us add, that when the physics constraints on
$m_q$ are relaxed and $m_q$ is set to 1~GeV, a good fit of the $\rho$
and $\phi$ data is obtained. Of course, this is only a hint that perhaps we
neglected in our analysis a mechanism that cuts off larger dipoles, e.g.\
vector meson size or saturation effects.
Data for $J/\psi$ have rather large errors and they are consistent with
zero. They are well described in the both approaches to the meson wave
function. If the data improved, though, the appearance of  deviations
from SCHC would indicate that one should go beyond the non-relativistic
approximation also in the $J/\psi$ case.

{\bf Energy dependence}. Measurements of $d\sigma/dt$ for $J/\psi$
photoproduction at $\gamma p$ energy 100~GeV and 200~GeV provide some
information on the value of the pomeron intercept. We used both LL BFKL
(with intercept $\alpha_P \simeq 1.45$) and a BFKL formalism modified
phenomenologically to incorporate non-leading corrections $\alpha_P \simeq
1.3$. The data show some growth with the energy which, within errors,
is consistent with both results, with slight preference for the lower
intercept.

\section{Conclusions}
Inclusion of BFKL evolution into a description of diffractive vector meson
photoproduction substantially improves the understanding of data. Having
very few free parameters, we get good fits to differential cross sections,
the correct hierarchy of helicity dependent amplitudes for $\rho$, $\phi$
and $J/\psi$ and the correct energy dependence for $J/\psi$
photoproduction. Only the sign of the single flip amplitude comes out
incorrect. This observable turns out, however, to be most sensitive to
contributions from colour dipoles of moderate size. Moreover, we confirmed
that the chiral odd components of the photon have to be taken into account 
in order to describe properly the light vector meson photoproduction.



\end{document}